\documentclass[apj]{emulateapj}

\usepackage{natbib}
\usepackage{color}
\usepackage{graphicx}
\usepackage{epstopdf}
\usepackage{amsmath}
\usepackage{booktabs}

\usepackage{float}
\usepackage[caption = false]{subfig}

\renewcommand{\deg}{\ensuremath{^{\circ}}}

\slugcomment{Accepted to ApJ on February 10, 2016}

\shorttitle{HAT-P-13b Core Mass}
\shortauthors{Buhler et al.}

\begin{document}


\title{Dynamical Constraints on the Core Mass of Hot Jupiter HAT-P-13b}


\author{Peter B. Buhler\altaffilmark{1}, Heather A. Knutson\altaffilmark{1}, Konstantin Batygin\altaffilmark{1}, Benjamin J. Fulton\altaffilmark{2}, Jonathan J. Fortney\altaffilmark{3}, Adam Burrows\altaffilmark{4}, Ian Wong\altaffilmark{1}}
\email{bpeter@caltech.edu}

\altaffiltext{1}{Division of Geological and Planetary Sciences, California Institute of Technology, Pasadena, CA, USA}
\altaffiltext{2}{NSF Graduate Research Fellow, Institute for Astronomy, University of Hawaii at Manoa, Honolulu, HI, USA}
\altaffiltext{3}{Department of Astronomy and Astrophysics, University of California, Santa Cruz, Santa Cruz, CA, USA}
\altaffiltext{4}{Astrophysical Sciences, Princeton University, Princeton, NJ, USA}


\begin{abstract}
HAT-P-13b is a Jupiter-mass transiting exoplanet that has settled onto a stable, short-period, and mildly eccentric orbit as a consequence of the action of tidal dissipation and perturbations from a second, highly eccentric, outer companion. Due to the special orbital configuration of the HAT-P-13 system, the magnitude of HAT-P-13b's eccentricity ($e_b$) is in part dictated by its Love number ($k_{2_b}$), which is in turn a proxy for the degree of central mass concentration in its interior. Thus, the measurement of $e_b$ constrains $k_{2_b}$ and allows us to place otherwise elusive constraints on the mass of HAT-P-13b's core ($M_{\rm{core,b}}$). In this study we derive new constraints on the value of $e_b$ by observing two secondary eclipses of HAT-P-13b with the Infrared Array Camera on board the $\textit{Spitzer Space Telescope}$. We fit the measured secondary eclipse times simultaneously with radial velocity measurements and find that $e_b = 0.00700 \pm 0.00100$. We then use octupole-order secular perturbation theory to find the corresponding $k_{2_b} = 0.31^{+0.08}_{-0.05}$. Applying structural evolution models, we then find, with 68\% confidence, that $M_{\rm{core,b}}$ is less than 25 Earth masses ($M_{\oplus}$). The most likely value of $M_{\rm{core,b}} = 11 M_{\oplus}$, which is similar to the core mass theoretically required for runaway gas accretion. This is the tightest constraint to date on the core mass of a hot Jupiter. Additionally, we find that the measured secondary eclipse depths, which are in the 3.6 $\mu$m and 4.5 $\mu$m bands, best match atmospheric model predictions with a dayside temperature inversion and relatively efficient day-night circulation.

\end{abstract}




\section{Introduction}
The interiors of gas giant planets provide ground truth for planet formation theories and the properties of materials under high pressure and temperature. Accordingly, many studies aimed at deriving the interior states of giant planets in our solar system have been undertaken in the past half century~\citep[e.g.,][]{Safronov1969,Mizuno1980,Stevenson1982,Bodenheimer1986,Pollack1996,Ikoma2000,Hubickyj2005,Rafikov2006,FortneyNettelmann2010,Nettelmann2012,Helled2013}. The study of giant planets in our solar system has been recently augmented by the growing body of mass and radius measurements for transiting extrasolar planets. These measurements have enabled the first studies of the heavy element components of gas giants orbiting other stars, as has been done for the super-Neptune HATS-7b~\citep{Bakos2015} and the hot Saturn HD 149026b~\citep{Sato2005}, and in the statistical characterization of heavy-element enrichment in extrasolar gas giant planets~\citep[e.g.,][]{Burrows2007,MillerFortney2011}. Nonetheless, characterizing the interior structure of exoplanets$-$in particular, determining the presence of a heavy element core$-$remains challenging, since mass and radius measurements alone cannot in general uniquely constrain the interior density profile nor the chemical makeup of a planet. In particular, determining whether heavy elements are concentrated in the core or distributed uniformly within the envelope is especially difficult for Jupiter-sized planets since the large, predominantly light-element envelope masks the signal of the radial distribution of heavy elements.

However, the orbital configuration in a subset of multi-planet systems is such that the dynamical evolution of the system depends on the Love number ($k_2$) of its innermost planet~\citep{Batygin2009}. The Love number ($k_2$) quantifies the elastic deformation response of a planet to external forces and thus encodes information about its interior structure, including clues about its core mass~\citep{Love1909,Love1911}. Utilizing the secular theory of~\citet{Mardling2007},~\citet{Batygin2009} showed that, in a system of two planets orbiting a central body, $k_2$ of the inner planet can be determined if (i) the mass of the inner planet is much smaller than the mass of the central body, (ii) the semi-major axis of the inner planet is much less than the semi-major axis of the outer planet, (iii) the eccentricity of the inner planet is much less than the eccentricity of the outer planet, (iv) the planet is transiting, and (v) the planet is sufficiently close to its host star, such that the tidal precession is significant compared to the precession induced by relativistic effects. The HAT-P-13 system is the first and only currently known system to fulfill these criteria.

The HAT-P-13 system consists of three bodies in orbit around a central star with a mass of $M_A=1.3\  M_{\odot}$ and radius $R_A=1.8\  R_{\odot}$~\citep{Southworth2012}. HAT-P-13b is a low-eccentricity transiting planet with mass $M_b=0.9\ M_J$, radius $R_b=1.5\ R_J$, and an orbital period of 2.9 days~\citep{Southworth2012}. HAT-P-13c is a radial velocity companion with a minimum mass $M_c = 14.2\ M_J$, an orbital period of 446 days, and an eccentricity of 0.66~\citep{Winn2010}. This system also exhibits a long-term radial velocity trend indicative of a third companion located between $12-37$ AU with a minimum mass of $15-200\ M_J$~\citep{Winn2010,Knutson2014}. However,~\citet{Becker2013} demonstrated that the existence of this third companion does not disrupt the secular dynamics that allows the eccentricity of HAT-P-13b ($e_b$) to be related to its Love number ($k_{2_b}$).

Using existing constraints on the orbital eccentricity of HAT-P-13b from radial velocity measurements,~\citet{Batygin2009} were able to place an upper bound on the core mass ($M_{\rm{core,b}}$) of 120 $M_{\oplus}$ ($41\%\  M_b$). In this study we present new observational measurements of secondary eclipses of HAT-P-13b (i.e. when HAT-P-13b passes behind its host star) obtained using $\textit{Spitzer Space Telescope}$, which we use to place stronger constraints on the eccentricity of HAT-P-13b. We combine these new secondary eclipse times with the most recent transit and radial velocity measurements of the system~\citep{Winn2010,Southworth2012,Knutson2014} in order to derive an improved constraint on $k_{2_b}$ and $M_{\rm{core,b}}$.

The paper is structured as follows. First, we describe our data acquisition, post-processing, and analysis (Section 2). We then present the results of the secondary eclipse measurements and corresponding determination of the eccentricity, $k_2$, core mass, and atmospheric properties of HAT-P-13b (Section 3).  Finally, we discuss the implications of our findings in Section 4.

\section{Methods}
\label{sec:methods}
\subsection{Observations and Photometric Time Series Extraction}
\label{subsec:aquisition}

Two observations of HAT-P-13 were taken using the InfraRed Array Camera (IRAC) on board the $\textit{Spitzer Space Telescope}$~\citep[$\textit{SST}$;][]{Fazio2004}, one using the 3.6 $\mu$m band on UT 2010 May 09 and the other using the 4.5 $\mu$m band on UT 2010 June 08, 11 orbits later (PI J. Harrington, Program ID 60003). Each dataset comprises 68,608 sub-array images taken with 0.4 s integration times over 8.7 hours of observation.

We extract the UTC-based Barycentric Julian Date (BJD$_{\rm{UTC}}$), subtract the sky background, and remove transient hot pixels from each of the images as described in~\citet{Knutson2012} and~\citet{Kammer2015}. To calculate the flux from the HAT-P-13 system in each image, we first estimate the position of the star on the array using the flux-weighted centroid method~\citep{Knutson2012,Kammer2015} with radii ranging between 2.0-5.0 pixels in 0.5 pixel increments. We then calculate the corresponding stellar flux using a circular aperture with either a fixed or time-varying radius.  We consider fixed radii ranging between 2.0-5.0 pixels in 0.5 pixel increments, and calculate the time-varying aperture using the square root of the noise pixel parameter as described in~\citet{Lewis2013}.  This parameter is proportional to the full width at half max (FWHM) of the star's point spread function, and is calculated for each image using a circular aperture with radii ranging between 2.0-5.0 pixels in 0.5 pixel increments.  We then either multiply the square root of the noise pixel parameter by a constant scaling value of [0.6, 0.7, 0.8, 0.85, 0.9, 0.95, 1.00, 1.05, 1.10, 1.15, or 1.20] pixels or add a constant offset of [-0.9, -0.8, -0.7, -0.6, -0.5, -0.4, -0.3, -0.2, -0.1, 0.0, 0.1, 0.2, 0.3, 0.4, or 0.5] pixels in order to determine the aperture radius for each image.

\subsection{Instrumental Noise Model and Optimal Aperture Selection}
\label{subsec:apselect}

We next create a time series for each photometric aperture where we trim the first 90 minutes (11,904 images) of each time series in order to remove the well-known ramp that occurs at the start of each new telescope pointing~\citep[e.g.,][]{Deming2006,Knutson2012,Lewis2013,Kammer2015}. We replace non-numerical (NaN) flux values with the median flux value of each time series and replace values that deviate by more than three standard deviations from the local mean, determined from the nearest 100 points, with the local mean. We compare this approach to one in which we instead trim outliers from our light curves and find that our best-fit eclipse depths and times change by less than 0.2 sigma in both channels. 0.2\% of the measurements were outliers or NaN in each channel. We then normalize each time series to one by dividing by the median value.

The photometric time series in both channels is dominated by an instrumental effect related to IRAC's well-known intrapixel sensitivity variations, combined with the pointing oscillation of the $\textit{SST}$.  We correct for this effect using Pixel Level Decorrelation (PLD), as described by~\citet{Deming2015}.  This method models the variation in flux intensity in each image due to this instrumental effect by tracking the change in intensity over time within a small box of pixels centered on the flux-weighted centroid. We use a total of nine pixels arranged in a 3x3 box centered on the position of the stellar centroid. We remove images from the time series where one of these nine pixels deviates from its mean flux by more than 3$\sigma$ (0.3\% of the data at 3.6 $\mu$m and 0.1\% of the data at 4.5 $\mu$m). Most of these deviations correlate with large pointing excursions in the photometric time series. We identify two pointing excursions in the 3.6 $\mu$m data, one of 0.7 pixels for 10 s and one of 0.5 pixels for 20 s, and one of 0.9 pixels for 10 s the 4.5 $\mu$m data. 

We divide the flux in each individual pixel by the summed flux across all nine pixels, weighting each pixel by its contribution to the flux and thereby isolating the instrument noise from astrophysical signals~\citep[see][]{Deming2015}, and repeat this operation for each image in our photometric time series. We also incorporate a constant and a linear term in time to model baseline instrument noise. Unlike~\citet{Deming2015}, we do not include a quadratic term because we found that the linear fit has an equivalent RMS (root mean square residual) to the quadratic fit and so adding the quadratic parameter is not justified. In addition, the quadratic term was correlated with the eclipse depth in our model fits.

We fit a combined instrumental noise and eclipse~\citep[][]{Mandel2002} model to the light curve for each combination of photometric apertures listed in Sec.~\ref{subsec:aquisition} using the `leastsq' routine in SciPy v0.14.0 with Python 2.7.6 and examine the residuals from the best-fit solution in order to determine the optimal aperture set for each bandpass. As discussed in~\citet{Deming2015} and~\citet{Kammer2015}, we first bin the photometric light curves and time series for individual pixels by a factor of 512 ($\sim$4 minute intervals) before fitting the model, then apply the resulting best-fit model coefficients to the unbinned light curve.  This allows us to identify solutions that minimize noise on longer time scales, which are most important for determining the best-fit eclipse parameters, in exchange for a moderately higher scatter in the unbinned residuals. We allow the center of eclipse time, eclipse depth, pixel weights, constant, and linear terms to vary as free parameters in our fits.

\begin{figure}
\epsscale{1.3}
\plotone{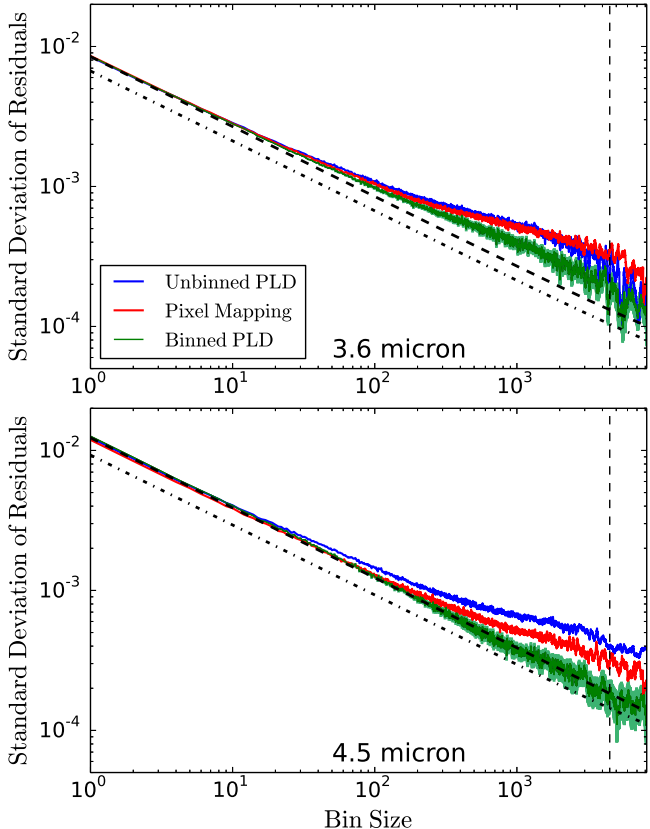}
\caption{The standard deviation of the residuals are normalized to match the standard deviation of the unbinned residuals for the PLD performed on data that was optimally binned before fitting (green), PLD that was not binned before fitting (blue), and the~\citet{Wong2014} pixel mapping fit (red) and plotted for each bandpass as a function of bin size. The vertical dashed line indicates the timescale of the eclipse ingress and egress. The expected $\sqrt{M/(n\times(M-1))}$ Gaussian scaling relation~\citep{Winn2008} of the standard deviation of the residuals as a function of the number of points per bin is also plotted (black dash-dot line is normalized to the Poisson noise and black dashed line is normalized to the standard deviation of the unbinned residuals for the PLD performed on data that was optimally binned before fitting; $M$ is the number of bins, $n$ is the bin size). The 1$\sigma$ uncertainties in the RMS ($RMS$/$\sqrt{2M}$) of the binned PLD model are plotted in light green.
\label{fig:RootN}}
\end{figure}

We excluded from consideration any apertures with an unbinned RMS more than 1.1 times that of the aperture with the lowest RMS in each band, focusing instead on the subset of apertures with low scatter. We then compared the relative amounts of time-correlated or ``red'' noise in the remaining apertures by calculating the standard deviation of the residuals as a function of bin size.  For light curves with minimal red noise, we would expect the standard deviation of the residuals to vary by the $\sqrt{M/(n\times(M-1))}$ Gaussian scaling relation~\citep{Winn2008}  where $n$ is the number of points in each bin and $M$ is the number of bins.  We evaluate the actual amount of red noise in the time series for each aperture by calculating the least-squares difference between the observed and theoretical noise scaling (Fig.~\ref{fig:RootN}) and select the aperture that minimizes this quantity in each bandpass.

We next find the optimal bin size to use to fit the lightcurve in each channel via the same least-squares approach with which we find the optimal aperture. After determining the optimal bin size in each bandpass, we repeat our aperture optimization at the new bin size. We iterate on searching for the optimal aperture and bin size until we converge on the optimal pairing of aperture and bin size for each bandpass.

After optimizing our choice of bin size and aperture, we found that the 4.5 $\mu$m light curve displayed a residual ramp-like signal despite our decision to trim the first 90 minutes of data. We therefore experimented with fits where we trimmed up to 3 hours of data from the start of the light curve (i.e., up to the beginning of the eclipse). We found that the best-fit eclipse times were correlated with the amount of data trimmed from the start of the light curve over the full range of trim durations considered, indicating that the ramp extended to the start of the eclipse.  We then considered an alternative approach in which we returned to our original 90 minute trim duration and deliberately used larger than optimal bin sizes in our fits, effectively forcing the models to identify solutions with less structure on long time scales.  We found that fits with bin sizes larger than 100 points (40 s) effectively removed the ramp from the light curve, avoiding the need to increase the trim interval to values larger than 90 minutes.  These fits resulted in best-fit secondary eclipse times approximately 2 minutes (0.6 $\sigma$) earlier than our original fits with a smaller bin size.  We tested for a residual ramp by repeating the large bin size fits with trim intervals ranging from 30 minutes up to 3 hours, and found no evidence for a correlation between the trim interval and the best-fit eclipse time.  We then repeated our optimization for bin size considering bin sizes between 128-2048 points in powers of two.  We found that our best-fit eclipse depths and times varied by less than 0.4$\sigma$ across this range, and were in good agreement with the best-fit values for the 3-hour trim interval using the smaller bin size. We also considered fits using a smaller bin size where we included an exponential function of time to account for the observed ramp, but found that this exponential function was a poor match for the shape of the observed ramp.  We speculate that a sum of several exponentials might provide a better fit~\citep[e.g.,][]{Agol2010}, but felt that the added free parameters were not justified given the success of using larger bin sizes. We also find that enforcing larger bin sizes in the 4.5 $\mu$m channel leads to better agreement of the secondary eclipse timing between the two channels.

We also tried decorrelating instrumental noise in our data using pixel-mapping \citep[e.g.,][]{Ballard2010,Lewis2013,Wong2014}. This non-parametric technique constructs an empirical map of the pixel response across the chip by comparing the measured flux from each image to those of other images with similar stellar positions. We model the pixel sensitivity at each point in our time series using a Gaussian spatial weighting function over the 50 nearest neighbors in stellar centroid $x$ and $y$ position and noise pixel parameter space. The inclusion of the noise pixel parameter in the weighting ensures that the pixel map incorporates systematics unrelated to changes in the star's position that affect the shape of the stellar point-spread function. The number of neighbors was chosen to be large enough to adequately map the pixel response across the range of star positions in each eclipse data set while maintaining a reasonably low computational overhead \citep{Lewis2013}.

\citet{Deming2015} found that PLD is generally more effective in removing time-correlated (i.e., red) noise than other decorrelation methods as long as the range of star positions across the data set remains below $\sim$0.2~pixels. The range of star centroid positions in our eclipse data sets lies below this threshold, and therefore, we expect PLD to perform optimally. We also directly compare the performance of PLD for cases where we fit to either the unbinned or optimally binned photometry, as well as to the fit acquired from photometry using the~\citet{Wong2014} pixel mapping technique described in the previous paragraph. We find that that the optimally binned PLD have lower levels of correlated noise than the other methods (Fig.~\ref{fig:RootN}). In addition, binned PLD gives center of eclipse phases in the two bandpasses that are most consistent with each other (at the 1.3$\sigma$ level); the unbinned PLD and pixel mapping techniques produced center of eclipse phases consistent at the 2.6$\sigma$ and 5.0$\sigma$ levels, respectively. We therefore select the PLD technique applied to the binned dataset for our final analysis.

For the fits described in the rest of this paper we use the following optimal aperture set and bin size. For the 3.6 $\mu$m channel we select a bin size of 21 points ($\sim$8 sec), a 3.0 pixel radius aperture to find the centroid, a 2.0 pixel aperture to find the noise pixel parameter, and add 0.3 pixels to the square root of the noise pixel parameter to obtain the aperture within which we sum the flux.  For the 4.5 $\mu$m channel we select a bin size of 128 points ($\sim$50 sec), a 4.5 pixel radius aperture to find the centroid, a 4.0 pixel aperture to find the noise pixel parameter, and add 0.3 pixels to the square root of the noise pixel parameter to obtain the aperture within which we sum the flux.

\begin{figure*}
\epsscale{1.0}
\subfloat{\includegraphics[width = 3.6in]{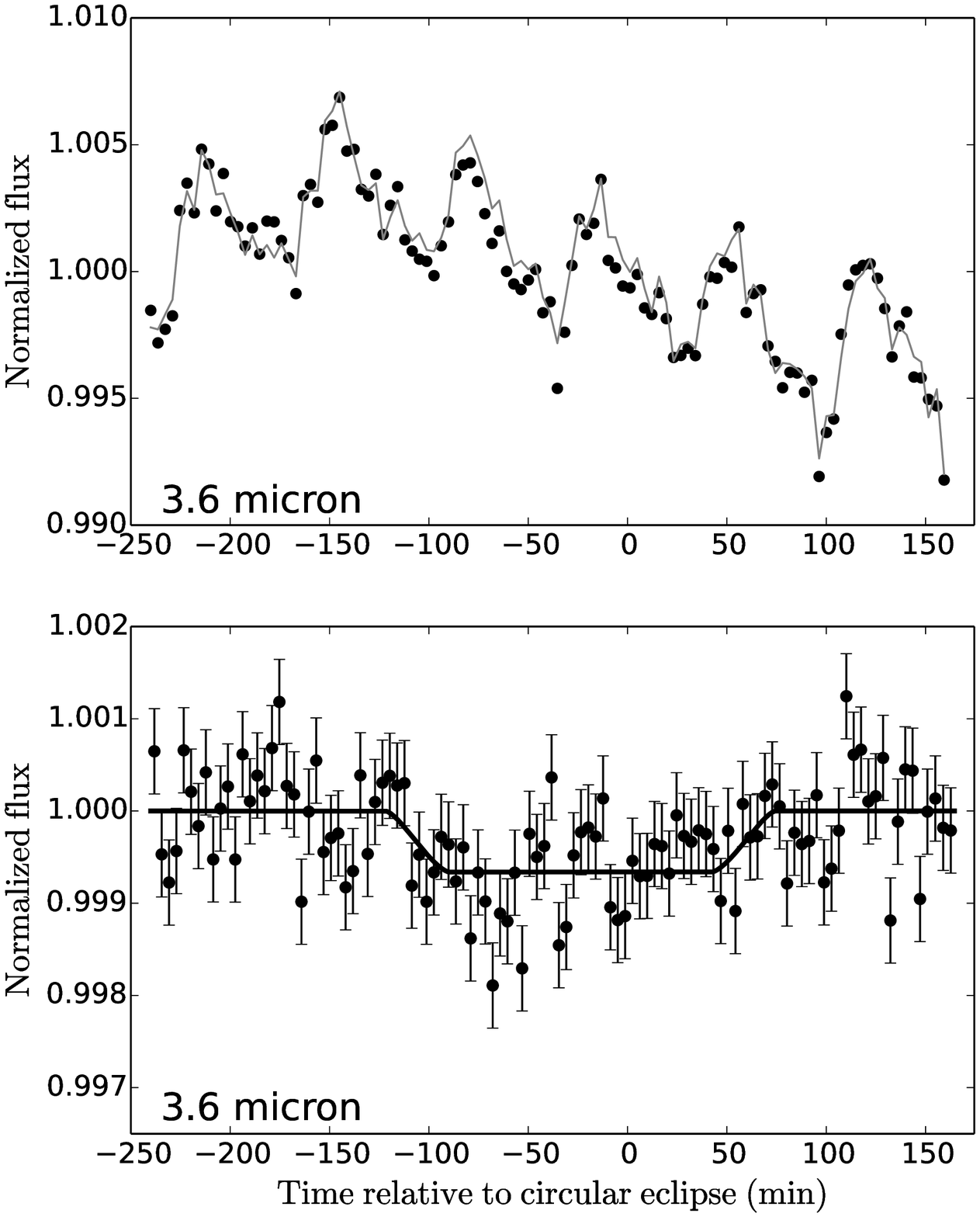}}
\subfloat{\includegraphics[width = 3.6in]{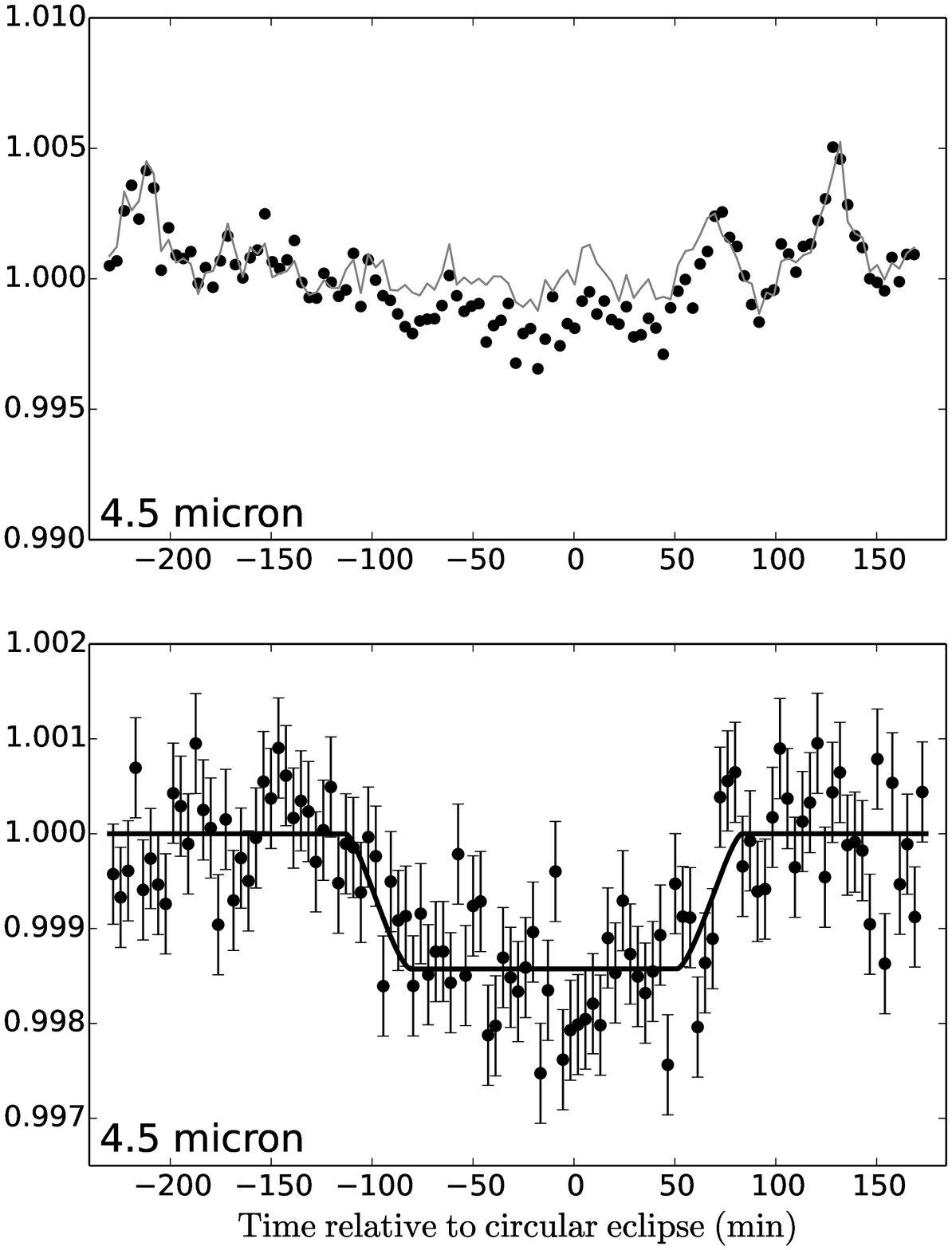}}
\caption{The top row shows the normalized raw flux (black points) compared to the best-fit instrumental noise model (gray line). The bottom row shows the best fit eclipse model (black line) and flux measurements after dividing out the instrumental noise model (black points). All data and models are plotted with a bin size of 512 measurements ($\sim$3.5 min) for visual clarity.
\label{fig:eclipses}}
\end{figure*}

\subsection{Eclipse Statistical Errors}
\label{subsec:eclipseerr}

We determine the uncertainties on our model parameters using the Markov-Chain Monte Carlo (MCMC) code emcee v2.1.0~\citep{ForemanMackey2013} on Python 2.7.6. We allow the center of eclipse time, eclipse depth, pixel weights, constant, and linear terms to vary as free parameters in our fits. We set the uncertainties on individual points in each light curve equal to the standard deviation of the residuals after subtracting the best-fit solution in each bandpass. We run the MCMC with 250 walkers for 20,000 steps; the first 5000 steps from each walker were `burn-in' steps and removed from the chain. 

For the observations in the 4.5 $\mu$m band we found that the 1$\sigma$ uncertainties on the RMS overlap with the errors theoretically expected in the absence of correlated noise on the timescale of the eclipse ingress and egress (30 min; Fig.~\ref{fig:RootN}) and therefore report the uncertainties in measurements from the 4.5 $\mu$m band directly from the MCMC analysis. However, for the observations in the 3.6 $\mu$m band, the calculated RMS consistently deviate above the expected improvement with increased binning for timescales longer than 1 min. We therefore choose a conservative approach and multiply the uncertainties in the center of eclipse time derived from the MCMC in the 3.6 $\mu$m band by a factor of 1.3, the factor by which the RMS lies above the theoretical improvement at the 30 min timescale~\citep{Pont2006,Winn2007}. Since the timescale of the eclipse is approximately half of the length of the dataset, we are unable to accurately estimate the red noise on that timescale and so adopt the same factor of 1.3 scaling for the eclipse depth uncertainty in this band.

\subsection{Eccentricity Determination}
\label{subsec:eccdet}

We next calculate an updated value for the eccentricity of HAT-P-13b using the approach described in~\citet{Fulton2013}.  We fit the available radial velocity observations for this planet from~\citet{Knutson2014} simultaneously with the best-fit transit ephemeris from~\citet{Southworth2012} and measured secondary eclipse times from this study. We first allow the apsides of each planet ($\omega_b$ and $\omega_c$) to vary independently and then repeat the fits imposing a prior that the posterior distribution of $\omega_b$ matches the posterior distribution of $\omega_c$ that was calculated from the fit in which $\omega_b$ and $\omega_c$ were allowed to vary independently.  We use the latter version of the fits in our final analysis, and discuss the rationale for this assumption in Sec.~\ref{subsec:seculardynamics} and~\ref{subsec:apsalignment}.

\subsection{Interior Modeling}
\label{subsec:interiormodel}

We use the MESA code~\citep{Paxton2011}, a 1-dimensional thermal evolution model, for interior modeling. In the pressure-temperature space relevant to HAT-P-13b, MESA uses the SCvH tables~\citep{Saumon1995} for the equation of state. We adopt a solar composition envelope and evolve an array of interior models of HAT-P-13b with varying core masses and energy dissipation rates. Specifically, we consider core masses of 0.1-80 $M_{\oplus}$ and dissipation rates equal to 0.05\%, 0.10\%, or 0.50\% of the insolation. The thermal dissipation range we adopt here encapsulates both (i) the energy deposition typically quoted for hot Jupiters residing on circular orbits (e.g. Ohmic dissipation, kinetic deposition) and (ii) an additional component of energy arising due to the sustained tidal dissipation~\citep[e.g.,][]{Bodenheimer2003,Batygin2009}. We calculate the insolation ($I$) using an equilibrium temperature of 1725 K~\citep{Southworth2012}.

We assume that the total mass of HAT-P-13b is the best fit value reported by~\citet{Winn2010}, 0.906 $M_J$, and acknowledge that a more recent value~\citep[0.899$M_J$,][]{Knutson2014} is available but that the mass-radius relationship for giant planets is famously independent of mass and so our choice of the~\citet{Winn2010} mass makes a negligible difference in our analysis. We also note that the errors on the mass are negligible compared to the uncertainties inherent in the equation of state~\citep[see][]{FortneyNettelmann2010}. We assume a Bond albedo of zero and a core density of 10 $g\ cm^{-3}$; varying the core density by 2 $g\ cm^{-3}$ has a negligible affect on the radial density profile obtained by MESA. We let the MESA models evolve for 3.0 Gyr, based on the best fit age of 3.5 Gyr reported by~\citet{Southworth2012}. However, the radial density structure reaches a quasi-steady solution after $\sim$1 Gyr, so the results are insensitive to the assumed system age.

For each pairing of core mass and dissipation rate we calculate $k_{2_b}$ based on the density profile, using the equations of~\citet{Sterne1939}\footnote[1]{Note that the definition of $k_{2,1}$ in~\citet{Sterne1939} is the \textit{apsidal motion constant}, i.e. $k_{2_b}$/2 in the notation used here.}:

\begin{equation}
\footnotesize
k_{2_b} = \frac{3 - \eta_2(R)}{2 + \eta_2(R)}
\label{eq:Sternek2}
\end{equation}
\\

$R$ is the radius of the planet and $\eta_2(R)$ is a dimensionless quantity that is obtained by integrating the ordinary differential equation radially in $\eta_2(r)$ outward from $\eta_2(0) = 0$:

\begin{equation}
\footnotesize
 r \frac{d\eta_2(r)}{dr} + \eta_2(r)^2 - \eta_2(r) - 6 + \frac{6\rho(r)}{\rho_m(r)}(\eta_2(r) + 1) = 0
\label{eq:Sterne_eta}
\end{equation}
\\

In the above expression, $\rho$ is the density obtained from the density distribution $\rho(r)$ output from MESA, and $\rho_m(r)$ is the mean density interior to $r$. Note that if the core density is constant then $\eta_2(r_{\rm{core}}) = 0$, where $r_{\rm{core}}$ is the core radius, ~\citep[i.e. $k_2$ is 3/2 for a body of constant density, e.g.,][]{Ragozzine2008}.

We use a linear spline to interpolate the coarse grid of $k_{2_b}$ and $R_b$ values, corresponding to various core mass and dissipation input pairings evolved in MESA, along both the core mass axis and the dissipation axis, and extend the grid from $0.1-80 M_{\oplus}$ to $0-80 M_{\oplus}$ with a linear extrapolation.

Once we determine the model values of $k_{2_b}$ and $R_b$ for each pair of core mass and dissipation, we evaluate the probability of each core mass and dissipation pairing, given the probability distributions of the measured values of $k_{2_b}$ and $R_b$ for the HAT-P-13 system. While the probability distribution for $R_b$ is measured from observation, the probability distribution of $k_{2_b}$ must be calculated. We describe this calculation below.

\begin{table}[h!]
\begin{center}
\caption{HAT-P-13 System Properties}
\label{tab:properties}
\begin{tabular}{ll}
\toprule
$e_b\ $ & $0.00700 \pm 0.00100$ \\
$e_c\ $ & $0.6554^{+0.0021}_{-0.0020}$ \\
$M_b\ (M_J)^2$ & $0.899^{+0.030}_{-0.029}$ \\
$M_c \sin(i_c)\ (M_J)^2$ & $14.61^{+0.46}_{-0.48}$ \\
$M_*\ (M_{\odot})^1$ & $1.320 \pm 0.062$ \\
$R_b\ (R_J)^1$ & $1.487 \pm 0.041$ \\
$R_*\ (R_{\odot})^1$ & $1.756 \pm 0.046$ \\
$T_b\ (\rm{day})^1$ & $2.9162383 \pm 0.0000022$ \\
$T_c\ (\rm{day})^2$ & $445.82 \pm 0.11$ \\
$a_b\ (\rm{AU})^1$ & $0.04383 \pm 0.00068$ \\
$\gamma\ (\rm{m\ s^{-1}})$ & $ -11.76^{+0.93}_{-0.9}$ \\
$\dot{\gamma}\ (\rm{m\ s^{-1}\ day^{-1}})$ & $ 0.0545\pm 0.0012$ \\
$\rm{jitter}\ (m\ s^{-1})$ & $ 4.7^{+0.48}_{-0.43}$ \\
3.6$\mu$m eclipse depth& 0.0662 $\pm$ 0.0113\%\\
3.6$\mu$m eclipse time ($BJD_{\rm{UTC}}$)& 2455326.70818 $\pm$ 0.00406\\
3.6$\mu$m eclipse offset (min)& -24.2 $\pm$ 5.8\\
3.6$\mu$m eclipse phase& 0.49424 $\pm$ 0.00139\\
4.5$\mu$m eclipse depth& 0.1426 $\pm$ 0.0130\%\\
4.5$\mu$m eclipse time ($BJD_{\rm{UTC}}$)& 2455355.87672 $\pm$ 0.00226\\
4.5$\mu$m eclipse offset (min)& -15.5 $\pm$ 3.3\\
4.5$\mu$m eclipse phase& 0.49633 $\pm$ 0.00079\\

\bottomrule

\end{tabular}
\\

$^1$~\citet{Southworth2012}, $^2$~\citet{Knutson2014}
\end{center}
\end{table}

\subsection{Secular Perturbation Theory}
\label{subsec:seculardynamics}

The octupole-order secular theory of ~\citet{Mardling2007}, augmented with a description of a tidally-facilitated apsidal advance~\citep{Ragozzine2008}, can be used to describe the non-Keplerian components of motion in the HAT-P-13 system and provides a method by which the relationship between $e_b$ and $k_{2_b}$ can be obtained~\citep{Batygin2009}. In the HAT-P-13 system, tidal dissipation quickly drains energy and acts to circularize the orbit of HAT-P-13b. However, the presence of the distant and highly eccentric HAT-P-13c acts to prevent complete circularization of the orbit of HAT-P-13b. Instead, the system tends towards a nearly elliptic equilibrium point, which acts as an attractor in phase space. As long as the orbits of HAT-P-13b and HAT-P-13c are coplanar, this minimization is achieved through aligning the apsides. Apsidal alignment is typically reached within $\sim$3 circularization timescales~\citep{Mardling2007}. However, once orbital equilibrium is achieved, both orbits decay slowly and the orbital configuration remains quasi-stable for the rest of the lifetime of the system.

In order to maintain alignment of the apsides, the apsidal precession of both HAT-P-13b and HAT-P-13c must be equal. That is:

\begin{equation}
\footnotesize
\dot{\varpi}_{c_{sec}} = \dot{\varpi}_{b_{sec}} + \dot{\varpi}_{b_{tid}} + \dot{\varpi}_{b_{GR}} + \dot{\varpi}_{b_{rot}}
\label{eq:apsides}
\end{equation}
\\

The secular apsidal precession of HAT-P-13c, $\dot{\varpi}_{c_{sec}}$, dominates all other contributions to the total apsidal precession of HAT-P-13c. The terms that dominate the apsidal precession of HAT-P-13b are the secular precession, $\dot{\varpi}_{b_{sec}}$, the tidal precession, $\dot{\varpi}_{b_{tid}}$, and general relativistic precession, $\dot{\varpi}_{b_{GR}}$. The minor effects due to rotational precession, $\dot{\varpi}_{b_{rot}}$, are also included but we neglect the negligible contribution to the apsidal precession from the stellar rotational bulge~\citep[e.g.,][]{Batygin2009}. The equations of apsidal precession are comprehensively discussed in~\citet{Ragozzine2008} and given here for convenience:

\footnotesize
\begin{multline}
\dot{\varpi}_{c_{sec}} = \frac{3}{4} n_c \left(\frac{M_b}{M_*}\right) \left(\frac{a_b}{a_c}\right)^2 \frac{1}{\left( 1 - e_c^2 \right)^2} \\ 
	\left[ 1 - \frac{5}{4} \left( \frac{a_b}{a_c} \right) \left( \frac{e_b}{e_c} \right) \frac{1 + 4 e_c^2}{1 - e_c^2} cos(\varpi_b - \varpi_c)\right]
\label{eq:varpi_c}
\end{multline}

\footnotesize
\begin{multline}
\dot{\varpi}_{b_{sec}} = \frac{3}{4} n_b \left(\frac{M_c}{M_*}\right) \left(\frac{a_b}{a_c}\right)^3 \frac{1}{\left( 1 - e_c^2 \right)^{3/2}} \\ 
	\left[ 1 - \frac{5}{4} \left( \frac{a_b}{a_c} \right) \left( \frac{e_c}{e_b} \right) \frac{cos(\varpi_b - \varpi_c)}{1-e_c^2} \right]
\label{eq:varpi_bsec}
\end{multline}

\footnotesize
\begin{multline}
\dot{\varpi}_{b_{tid}} = \frac{15}{2} k_{2_b}  n_b \left(\frac{R_b}{a_b}\right)^{5} \left(\frac{M_*}{M_b}\right) \left( 1 - e_b^2 \right)^{-5} \\
\left( 1+ \frac{3}{2}e_b^2 + \frac{1}{8} e_b^4 \right)
\label{eq:varpi_btid}
\end{multline}

\begin{equation}
\footnotesize
\dot{\varpi}_{b_{GR}} = \frac{3n_b^3}{1 - e_b^2} \left(\frac{a_b}{c}\right)^2
\label{eq:varpi_GR}
\end{equation}

\begin{equation}
\footnotesize
\dot{\varpi}_{b_{rot}} = \frac{k_{2_b}}{2} \left(\frac{R_b}{a_b}\right)^5 \frac{n_b^3 a_b^3}{G m_b \left( 1 - e_b^2 \right)^2}
\label{eq:varpi_GR}
\end{equation}

\normalsize

In the preceding equations, $G$ is the Newtonian gravitational constant and $c$ is the speed of light. The subscripts `b', `c', and `*' denote properties of HAT-P-13b, HAT-P-13c, and the star, respectively. $a$ is the semimajor axis, $e$ is the eccentricity, $n$ is the mean motion, $R$ is the radius, and $M$ is the mass. Under the assumption that the apsides are aligned, the $\varpi_b-\varpi_c$ terms in Eqs.~\ref{eq:varpi_c} and~\ref{eq:varpi_bsec} are zero. Since all of the system properties that appear in the equations of apsidal precession have been measured, with the exception of $k_{2_b}$, Eq.~\ref{eq:apsides} can be rearranged to solve for the Love number of HAT-P-13b purely in terms of known quantities. Note that it is not necessary to measure the apsidal precession rate of either HAT-P-13b nor HAT-P-13c, it is sufficient to know only that they are equal.

\subsection{Core Mass Determination}
\label{subsec:coremass}

We construct the posterior probability distribution for $k_{2_b}$ from MCMC chains comprising 10$^7$ normally distributed values for each of the measured HAT-P-13 system properties (Table~\ref{tab:properties}) using Eqs.~\ref{eq:apsides}-\ref{eq:varpi_GR}. We then multiply the probability distributions for $k_{2_b}$ and $R_b$ obtained from MESA and map that distribution into a 2-dimensional probability distribution of core mass and heat dissipation. Finally, we obtain the 1-dimensional probability distribution of the core mass of HAT-P-13b by marginalizing the 2-dimensional distribution over dissipation, assuming a uniform prior on dissipation between 0.05\%-0.5\% $I$.

\subsection{Atmospheric Measurements}
\label{subsec:tempmeasure}

We determine the dayside temperature of HAT-P-13b from the measured secondary eclipse depths in each bandpass. To do so, we first calculate the stellar flux by integrating a PHOENIX stellar flux model~\citep{Husser2013} for each bandpass weighted by the subarray average spectral response curve\footnote[2]{Curve obtained from `Spectral Response' at \\ http://irsa.ipac.caltech.edu/data/SPITZER/docs/irac/calibrationfiles}. We utilize a PHOENIX model with an effective temperature of $T_\mathrm{eff}$ = 5700 K, a surface gravity of $\log{g}$ = 4.0, and a modestly enhanced metallicity of $[\mbox{Fe}/\mbox{H}]$ = 0.5. For comparison, HAT-P-13 has a measured $T_\mathrm{eff}$ = $5720\pm 69$ K, $[\mbox{Fe}/\mbox{H}]$ = $0.46\pm 0.07$~\citep{Torres2012}, and $\log{g}$ = $4.070\pm 0.020$~\citep{Southworth2012}. We calculate the flux of the planet as a fraction of the total system flux based on the depth of the secondary eclipse. We then find the temperature that gives a blackbody curve that, when integrated over its respective bandpass, matches the planetary flux. We calculate the errors on the temperature by constructing the posterior distribution for the temperature in each wavelength using MCMC chains of length 2.5 $\times$ 10$^4$, based on the measured eclipse depths and $R_b/R_*$. The effective dayside temperature was calculated by taking the error-weighted mean of the best-fit temperatures in each bandpass.

\section{Results}
\label{sec:results}

\begin{figure}
\epsscale{1.3}
\plotone{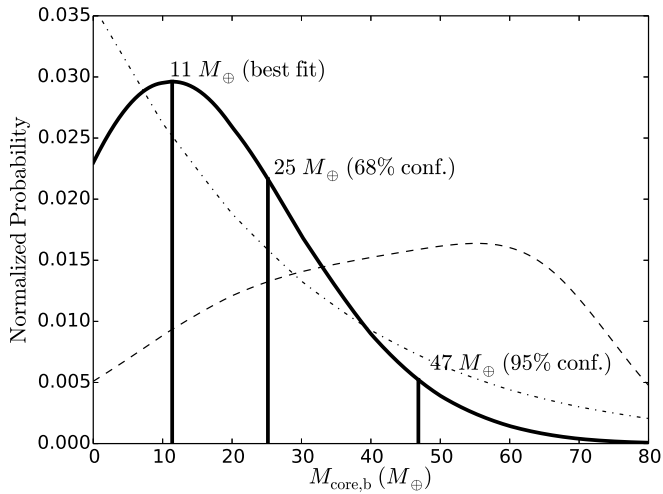}
\caption{The probability distribution of the true core mass of HAT-P-13b (black), along with the most probable core mass ($11\ M_{\oplus}$), 68\% confidence interval ($0-25\  M_{\oplus}$), and 95\% confidence interval ($0-47\ M_{\oplus}$) are shown. The probability distribution of the core mass is the product of the constraints on the core mass probability given by the measurement uncertainty in the Love number ($k_{2_b}$, dash-dot) and the radius ($R_b$, dashed).
\label{fig:CoreMass}}
\end{figure}

\subsection{Secondary Eclipse Measurements}
\label{sec:measurements}

We find that the HAT-P-13b secondary eclipses are centered at 2455326.70818 $\pm$ 0.00406 and 2455355.87672 $\pm$ 0.00226 BJD$_{\rm{UTC}}$ in the 3.6 $\mu$m and 4.5 $\mu$m bands, respectively. These times are 24.2 $\pm$ 5.8 minutes and 15.5 $\pm$ 3.3 minutes earlier (orbital phase 0.49424 $\pm$ 0.00139 and 0.49633 $\pm$ 0.00079), respectively, than the predicted time based on a circular orbit (Fig.~\ref{fig:eclipses}), where we have accounted for the 41 s light travel time delay~\citep{Loeb2005} and the uncertainty in the~\citet{Southworth2012} ephemeris (9.7 and 11 seconds for the 3.6 $\mu$m and 4.5 $\mu$m observations, respectively). The eclipse depths for the 3.6 $\mu$m and 4.5 $\mu$m channel are 0.0662 $\pm$ 0.0113\% and 0.1426 $\pm$ 0.0130\%, respectively (Fig.~\ref{fig:eclipses}).

These secondary eclipse times are consistent at the 1.3$\sigma$ level. We therefore take the error-weighted mean and find that the observed center of secondary eclipse time occurs 17.6 $\pm$ 2.9 minutes earlier (orbital phase 0.49582 $\pm$ 0.00069) than the predicted value for a circular orbit.

\begin{figure*}
\epsscale{1.0}
\subfloat{\includegraphics[width = 3.6in]{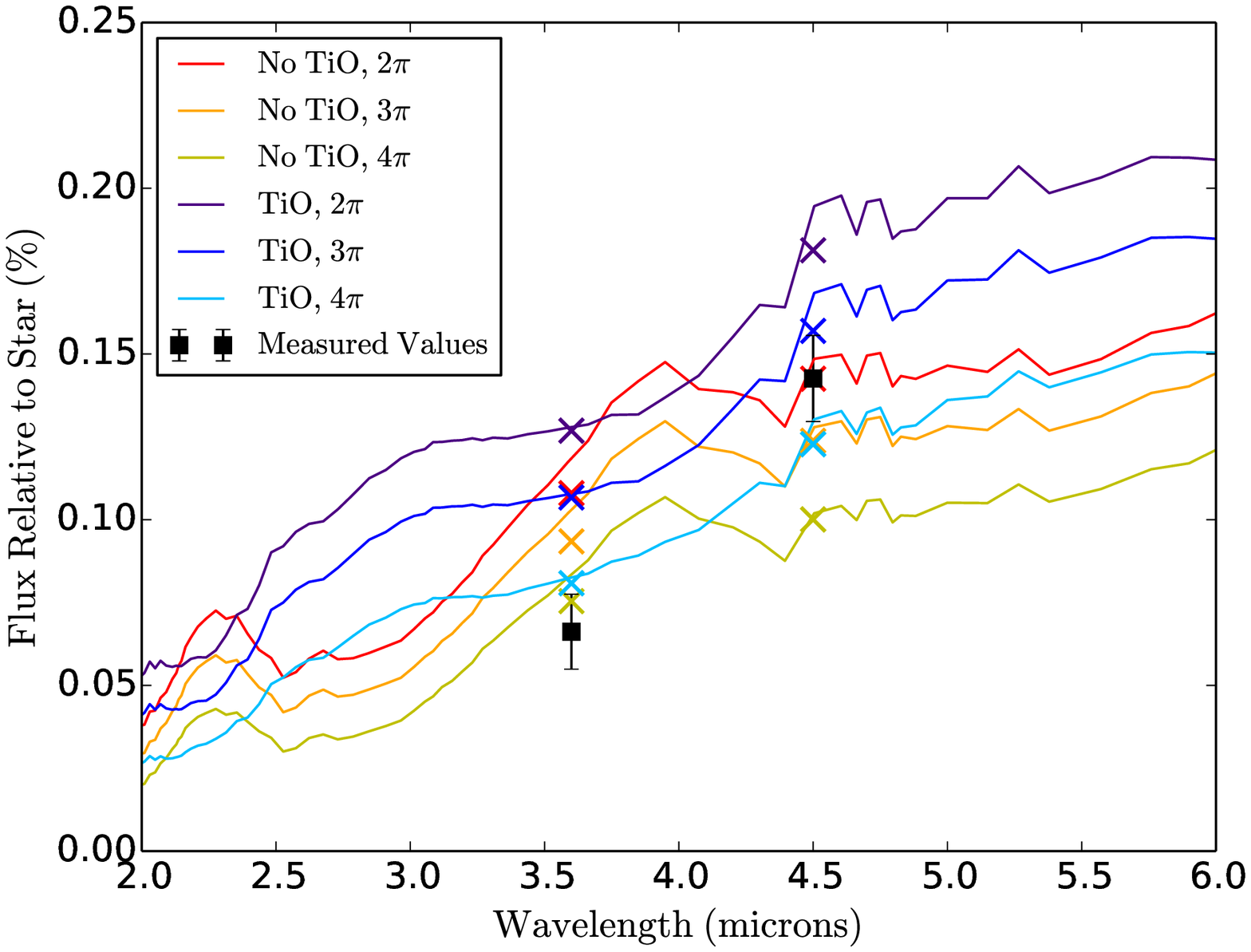}}
\subfloat{\includegraphics[width = 3.6in]{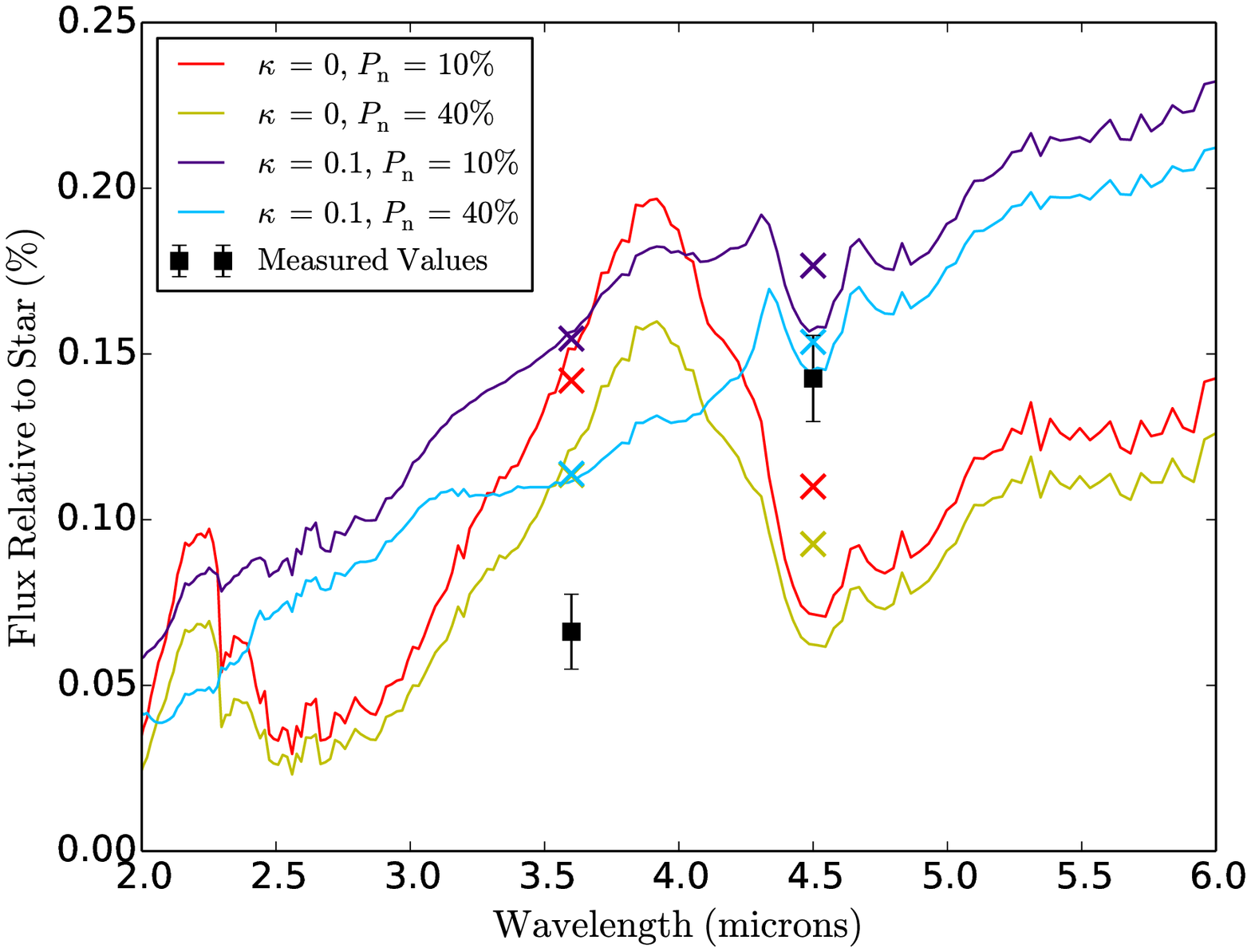}}
\caption{The left panel shows six dayside atmosphere models for HAT-P-13b based on~\citet{Fortney2008} and the right panel shows four models based on~\citet{Burrows2008}. The measured secondary eclipse depths at 3.6 and 4.5 microns are overplotted as black filled squares, and the band-integrated model predictions are shown as colored X's for comparison.~\citet{Fortney2008} models an atmospheric absorber with TiO and either no circulation (2$\pi$), partial circulation (3$\pi$), or full circulation (4$\pi$).~\citet{Burrows2008} models opacity with a gray source ($\kappa$, units of $cm^2g^{-1}$) and the fraction of energy redistributed to the night side ($P_n$; 10\% is minimal redistribution, 40\% is near-maximal redistribution).
\label{fig:atmosphere}}
\end{figure*}

\subsection{Eccentricity and Core Mass}
\label{sec:eccentricity}

Assuming apsidal alignment, the eccentricities of the orbits for the two innermost planets in this system are $e_b = 0.00700\pm 0.00100$ and $e_c = 0.6554_{-0.0020}^{+0.0021}$. We use these eccentricities to calculate a Love number for the innermost planet ($k_{2_b})$ of $0.31_{-0.05}^{+0.11}$, where values of $k_{2_b} > 0.30$ are inconsistent with the MESA interior models (i.e., would require a negative core mass). When we combine this constraint on $k_{2_b}$ with the measured planet radius ($R_b$) we find that the core mass of HAT-P-13b is less than 25 $M_{\oplus}$ (less than 9\% $M_b$; 68\% confidence interval), with a most likely core mass of 11 $M_{\oplus}$ (4\% $M_b$; Fig.~\ref{fig:CoreMass}).  The constraint from $k_{2_b}$ strongly favors  smaller core masses, while the constraint from $R_b$ modestly favors larger core masses, up to $\sim 60\ M_{\oplus}$ (Fig.~\ref{fig:CoreMass}).

\subsection{Atmospheric Properties}
\label{sec:temperature}

We find best-fit brightness temperatures of 1680 $\pm$ 119 K at 3.6 $\mu$m and 2265 $\pm$ 150 K at 4.5 $\mu$m and compare our measured eclipse depths in each bandpass to atmosphere models from~\citet{Burrows2008} and~\citet{Fortney2008} (Fig.~\ref{fig:atmosphere}). Both models assume a solar composition, plane-parallel atmosphere with molecular abundances set to the local thermal equilibrium values. The~\citet{Fortney2008} models assume even heat distribution across the dayside and vary the amount of energy incident at the top of the dayside atmosphere in order to approximate the effects of redistribution to the night side.  In these models the zero redistribution case is labeled as `2$\pi$' and the full redistribution case is labeled as `4$\pi$'.  We also consider versions of the model with and without an equilibrium abundance of TiO; when present, this molecule absorbs at high altitudes and produces a temperature inversion in the dayside atmosphere. The~\citet{Burrows2008} models account for the presence or absence of a dayside temperature inversion by introducing a gray absorber at low pressures where the opacity $\kappa$ can be adjusted as a free parameter.  Atmospheric circulation is included as a heat sink between 0.01-0.1 bars, where the parameter $P_n$ defines the fractional amount of energy redistributed to the night side and ranges from 0-50\% (from no redistribution to the nightside to complete redistribution across both hemispheres). The~\citet{Fortney2008} model satisfactorily reproduces the observed eclipse depths in both bandpasses when including a dayside temperature inversion due to absorption from TiO and relatively efficient circulation between the day and night sides. Although none of the~\citet{Burrows2008} models are able to match the observed 3.6 micron eclipse depth within the 3$\sigma$ uncertainty, we obtain the closest match with models that include an absorber ($\kappa$ = 0.1) and relatively efficient circulation ($P_n$ = 40\%).

\section{Discussion}
\label{sec:discussion}

\subsection{Effects of Coplanarity and Apsidal Alignment}
\label{subsec:apsalignment}

Correlations between the apsidal orientation ($\omega$) and eccentricity ($e$) introduce errors on the determination of eccentricity of HAT-P-13b ($e_b$). Since $e_b$ is relatively small, we obtain a correspondingly poor constraint on $\omega_b$ of $231_{-42}^{+17}$ degrees in fits where we allow $\omega_b$ to vary independently of $\omega_c$. However, since $e_c$ is large, we are able to measure $\omega_c$ with an uncertainty of less than a degree ($\omega_c = 175.28\deg$$^{+0.21}_{-0.22}$). The measured apsidal angles for planets b and c are thus consistent with apsidal alignment, although the relatively large uncertainties on $\omega_b$ preclude a definitive determination. 

When we allow $\omega_b$ to vary freely in our fits we find that $e_b = 0.0108_{-0.0035}^{+0.0069}$. This eccentricity is nonzero at the 3.1$\sigma$ level, providing independent confirmation that the orbit of HAT-P-13b has not yet been circularized and therefore that the secular orbital coupling mechanism discussed by~\citet{Mardling2007} and~\citet{Batygin2009} is applicable to this system. Note that the uncertainty in $e_b$ is more than five times greater than in the case when we assume apsidal alignment.

If the planets are coplanar, their apsides will align in much less than the age of the HAT-P-13 system~\citep{Mardling2007,Batygin2009}.~\citet{Mardling2010} showed that an initial mutual inclination between the orbits of HAT-P-13b and HAT-P-13c would evolve to a limit cycle in $e_b$ and apsidal orientation, rather than to a fixed $e_b$ and apsidal alignment. That study explored the effects of the inclination angle between the orbits of HAT-P-13b and HAT-P-13c ($\Delta i_{b-c}$) on $e_b$ and found that if the orbits are nearly coplanar ($\Delta i_{b-c} \leq 10\deg$) then the limit cycle in $e_b$ will have a width of less than $3\% \ e_b$ and the width of the limit cycle of the angle between the apsides is $\lesssim 4\deg$~\citep[calculated from Eq. 15, 16, and 17 of][]{Mardling2010}. Thus, the $e_b$ measured at a particular epoch of the HAT-P-13 system is insensitive to this limit cycle if $\Delta i_{b-c}$ is low.

We propose that $\Delta i_{b-c}$ is indeed likely to be small, based on both observational constraints and theoretical arguments. First, the exploration by~\citet{Mardling2010} found that a configuration of either (i) prograde, near-coplanar orbits or (ii) $130\deg \lesssim \Delta i_{b-c} \lesssim 135 \deg$ is strongly favored. Second,~\citet{Winn2010} measured the Rossiter-McLaughlin effect~\citep{Rossiter1924,McLaughlin1924} during a transit of HAT-P-13b and found that the spin axis of the star and the angular momentum vector of HAT-P-13b's orbit are well aligned on the sky ($\lambda = 1.9 \pm 8.6\deg$). This is significant because HAT-P-13b orbits far enough from the star that the the orbital precession rate is dominated by torque from HAT-P-13c rather than the $J_2$ quadrupole moment of the star~\citep{Mardling2010,Winn2010}. If $\Delta i_{b-c}$ were large, as in case (ii) of~\citet{Mardling2010}, nodal precession of HAT-P-13b's orbit around HAT-P-13c's orbital axis would ensue, manifesting as cyclic variations in the angle between stellar equator and the orbital plane of HAT-P-13b ($\psi_{*,b}$). Therefore, it is unlikely that a small value for  $\psi_{*,b}$ would be measured at a randomly selected epoch unless $\Delta i_{b-c}$ is small~\citep{Winn2010}. However, the initial orbital configuration of the system is unknown and the sky-projected angle ($\lambda$), rather than the true $\psi_{*,b}$, is measured, so it is not possible to definitively determine $\Delta i_{b-c}$ from the Rossiter-McLaughlin measurement alone. We therefore argue that $\Delta i_{b-c}$ must be small, without attempting to place a definitive upper limit on $\Delta i_{b-c}$.

A direct measurement of $\Delta i_{b-c}$ may be forthcoming by studying transit timing variations (TTVs) in the orbit of HAT-P-13b, since mutual inclination can induce a detectable TTV signature~\citep{Nesvorny2009}.~\citet{Southworth2012} found that there is no compelling evidence for large TTVs in the orbit of HAT-P-13b, although TTVs of less than 100 s are possible~\citep{Fulton2011}.~\citet{PayneFord2011} explored theoretical TTVs for HAT-P-13b and found that HAT-P-13c should induce TTVs on the order of tens of seconds, and that a precise determination of TTVs would make it possible to discriminate between the two allowed scenarios ($\Delta i_{b-c}$ near 0$\deg$ or 130-135$\deg$) found by~\citet{Mardling2010}. 

Astrometry of HAT-P-13 could also be used to probe $\Delta i_{b-c}$. We calculate an expected astrometric signal from HAT-P-13b of either (i) 61 $\mu$as, if the orbit of HAT-P-13c is effectively edge-on as seen from Earth or (ii) 86 $\mu$as if it is inclined at 135$\deg$ as seen from the Earth. Astrometry from the Gaia mission should be accurate to $\sim$10 $\mu$as~\citep{Lindegren2009} and thus will be sensitive enough to discriminate between these two scenarios. Although a direct measurement of the apsidal precession of the system (i.e., $\dot{\varpi_c}$) would allow a direct check of the secular perturbation theory that allows us to calculate $k_{2_b}$, we calculate that the precession rate for this planet is on the order of $10^{-4}$ deg/year and is therefore beyond the reach of current radial velocity observations. However, the presence of a third companion~\citep{Winn2010,Knutson2014} in the system may complicate the determination of $\Delta i_{b-c}$ using any of these methods.

\subsection{Interior Structure}

\begin{figure}
\epsscale{1.3}
\plotone{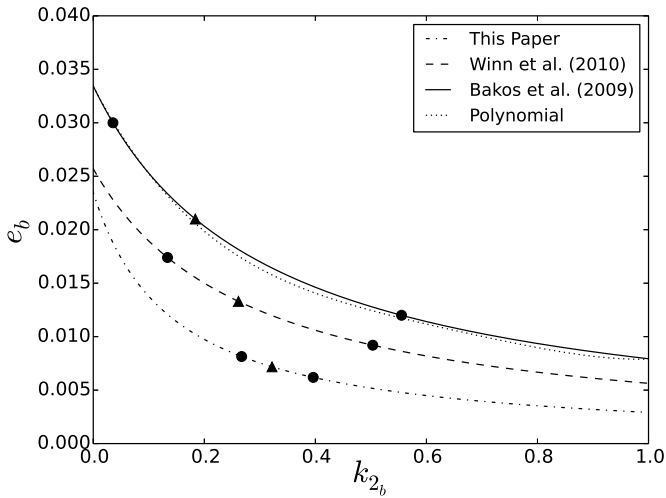}
\caption{The relationship between $e_b$ and $k_{2_b}$ for the HAT-P-13 system parameters measured by different studies, including the fourth order polynomial approximation given in~\citet{Batygin2009}. The best-fit (triangles) and $1\sigma$ (circles) uncertainties in $e_b$ reported by each study are plotted on their respective $e_b$-$k_{2_b}$ curves. The curves do not include uncertainties in the $e_b$-$k_{2_b}$ relationship due to measurement errors, unlike our Bayesian model (Fig.~\ref{fig:CoreMass}), which does take them into account.
\label{fig:k2_vs_e}}
\end{figure}

The initial characterization of $M_{\rm{core,b}}$ by~\citet{Batygin2009} was limited by the relatively large uncertainty in the published eccentricity for the innermost planet. Based on radial velocity data alone they concluded that $M_{\rm{core,b}}$ must be less than 120 $M_{\oplus}$ at the 1$\sigma$ level, and argued that core masses greater than 40 $M_{\oplus}$ were disfavored based on the required effective tidal dissipation ($Q_b$)~\footnote[3]{Although, their model did not account for other sources of heating such as Ohmic dissipation.}. More recently,~\citet{Kramm2011} used updated measurements of the HAT-P-13 system from~\citet{Winn2010} to find an allowed range of $k_{2_b}$ based on the 1$\sigma$ error on $e_b$ by using the polynomial relating $e_b$ and $k_{2_b}$ given in~\citet{Batygin2009}. They then used that $k_{2_b}$ range to place constraints on the interior structure of HAT-P-13b using the values of $M_b$ and $R_b$ from~\citet{Bakos2009} and complex interior models. Their analysis indicated that $M_{\rm{core,b}}$ is less than $27 M_{\oplus}$. However, caution must be exercised when using the polynomial equation of~\citet{Batygin2009}, since the shape of the curve strongly depends on all of the measured system parameters (Fig.~\ref{fig:k2_vs_e}). In addition, the polynomial does not include uncertainties in the $e_b$-$k_{2_b}$ relationship due to observational measurement uncertainties.

Our analysis offers an improved estimate of $M_{\rm{core,b}}$ (less than $25 M_{\oplus}$ with 68\% confidence) by taking into account both the change in the dependence of $k_{2_b}$ on $e_b$ due to updated measurements of $M_b$, $M_c$, $M_*$, $R_b$, $T_b$, $T_c$, and $e_c$ and the effect of the uncertainties in those measured values on the $e_b$-$k_{2_b}$ relationship and $M_{\rm{core,b}}$ determination, which had been neglected in previous studies. When combined with new radial velocity measurements from~\citet{Knutson2014}, the secondary eclipse measurements of HAT-P-13 provide strong constraints on $e_b$ and our assumption of apsidal alignment further reduces uncertainty on this parameter. Our method also allows us to explore the full probability distribution for $M_{\rm{core,b}}$ instead of only placing an upper bound on its value.

There are several caveats worth mentioning in regard to our estimated core mass.  We note that $k_2$ is only the lowest harmonic describing the internal yielding of a body to external forces, and is thus an inherently degenerate quantity~\citep[as noted for specific models of HAT-P-13b by][]{Kramm2011}. The effects of metallicity on atmospheric opacity may also affect the thermal evolution and thus the radial structure of the planet~\citep[as noted for brown dwarfs by][]{Burrows2011} but are neglected here. We adopt a solar composition envelope for definitiveness and expect that increasing the metallicity will have only a small effect on our predicted core mass based on the extensive exploration of this effect on interior models performed by~\citet{Kramm2011}. We also note that an inhomogeneous heavy element distribution may lead to an overestimation of $M_{\rm{core,b}}$~\citep{Leconte2012}. Thus, our estimate is specific to a model with a refractory element core and a solar composition envelope. Imperfect knowledge of the equations of state of materials at high pressure and temperature also introduces additional uncertainties~\citep[e.g.][]{FortneyNettelmann2010} that are not accounted for in this study.

In addition, strong constraints on the internal heat dissipation are not available, although we can determine how the uncertainty in the internal dissipation impacts our conclusions for $M_{\rm{core,b}}$ by re-calculating the $M_{\rm{core,b}}$ probability distribution assuming either extremely high or extremely low dissipation rates. We find that the main effect of the dissipation rate is to shift the peak of the probability distribution for $M_{\rm{core,b}}$ lower for higher values of dissipation, while maintaining a comparable distribution shape. When we specify dissipation as 0.05\% $I$, the probability distribution peaks at $M_{\rm{core,b}}$ = 22 $M_{\oplus}$. For a dissipation of 0.50\% $I$, the probability distribution peaks at $M_{\rm{core,b}}$ = 3 $M_{\oplus}$. We therefore conclude that uncertainties in the internal heat dissipation introduce modest, but not overwhelming, uncertainties in the estimate of $M_{\rm{core,b}}$ (i.e., lack of knowledge of the heat dissipation yields uncertainties that are within the 1$\sigma$ errors from the observational uncertainties).

\subsection{Dayside Atmosphere}

~\citet{Schwartz2015} compare the irradiation temperatures ($T_0 = T_*\sqrt{R_*/a_b}$) of a large sample of hot Jupiters to their measured dayside brightness temperatures ($T_d$) from secondary eclipse observations, and find that hotter planets appear to have relatively inefficient day-night circulation. For HAT-P-13b $T_0 = 2469$ K, yielding a predicted $T_d\approx 2090$ K~\citep[from Fig. 2 of][]{Schwartz2015}, which is 2$\sigma$ above the effective dayside temperature we measure ($1906 \pm 93$ K). The $T_d/T_0$ that we obtain for HAT-P-13b ($0.7720\pm 0.0377$) indicates relatively efficient redistribution of energy to the night side for the case of zero Bond albedo~\citep[see Fig. 7 of][]{Cowan2011}, in good agreement with our findings in Sec.~\ref{sec:temperature}. The $T_0/T_d$ of HAT-P-13b also fits the trend of decreasing $T_0/T_d$ with lower planetary mass found by~\citet{Kammer2015} (their Fig.~13). The circulation model of~\citet{Perez-Becker2013}, which depends on the equilibrium temperature of the planet, also predicts moderately efficient energy redistribution such that the nightside flux from HAT-P-13b should be 0.55-0.75 that of its dayside flux, depending on the drag timescale.

\subsection{Comparison to Other Systems}

Our analysis indicates that $M_{\rm{core,b}}$ is comparable to the core masses of Jupiter~\citep[$M_{\rm{core,J}} < 18 M_{\oplus}$,][]{FortneyNettelmann2010} and Saturn~\citep[$M_{\rm{core,S}} = 5$-$20 M_{\oplus}$,][]{Helled2013} in our own solar system. Core accretion models for gas giant planet formation suggest that minimum core masses of approximately 10 $M_{\oplus}$ are needed in order to form Jovian planets, although this limit depends on both the composition of the core and the properties of the gas disk near the planet's formation location~\citep[e.g][]{Mizuno1980,Bodenheimer1986,Pollack1996,Ikoma2000,Hubickyj2005,Rafikov2006}. Although our observation is consistent with core accretion theory~\citep{Safronov1969,Stevenson1982}, our 1$\sigma$ confidence interval extends down to zero core mass and therefore does not preclude alternative formation models such as disk instability~\citep[e.g.,][]{Boss1997}, nor does it provide a definitive test of post-formation core erosion~\citep[e.g.][]{Stevenson1982, Guillot2004}.

Work has been undertaken to probe the heavy-element fractions of gas giant planets across a broad range of planets, from the hot super-Neptune HATS-7b~\citep{Bakos2015} and hot Saturn HD 149026b~\citep[e.g.,][]{Sato2005,Ikoma2006,Fortney2006,Burrows2007,Southworth2010} to super-Jupiters~\citep[e.g. GJ436b and HAT-P-2b;][]{Baraffe2008}. The constraints on the heavy-element component of these planets are often accompanied by statements about their inferred core mass, with the caveat that there are degeneracies between models with heavy element cores and models with heavy elements distributed throughout the envelope~\citep[e.g.,][]{Baraffe2008}. Avenues for partially breaking the degeneracies between thermal evolutionary models with heavy-elements distributed throughout the planet and models with heavy-element cores are available for extremely metal-rich planets, such as HATS-7b and HD 149026b. However, in general, measurements of mass and radii can only be used to constrain the overall fraction of the planetary mass composed of heavy elements. The inference of a radial distribution of refractory elements, and therefore assertions related to the mass of a solid core, require additional information (e.g. knowledge of $k_2$). In this regard HAT-P-13b is unique because it is the only member of the extrasolar planetary census for which this additional information exists. 

Our constraint on the core mass of  HAT-P-13b is consistent with the determination of heavy-element enrichment, with the accompanying inference of the presence of cores in hot Jupiters by~\citet{Torres2007} and~\citet{Burrows2007}.~\citet{Torres2007} invoke the presence of heavy element cores to explain the small radii of the metal-rich 0.60 $M_J$ HAT-P-3b and 0.62 $M_J$ XO-2b, and~\citet{Burrows2007} investigated a sample of 14 hot Jupiters and found that a subset of those planets had smaller radii than allowed by models without either a solid core or metal-rich envelope. We stress, though, that the independent measurement of the degree of central mass concentration, such as done in this paper, is necessary to determine the radial distribution of heavy elements for Jovian-mass planets.

Finally, we also compare the results of our study to empirical scaling relations from~\citet{MillerFortney2011}, which are based on mass and radius measurements from a sample of 15 planets with moderate irradiation levels (incident flux $< 2 \times 10^8$ erg s$^{-1}$) around stars with metallicities ranging from [Fe/H]$_*$ = -0.030 to +0.390. That study found a positive correlation between the bulk metallicity of a planet and that of its host star and a negative correlation between a planet's mass and its metallicity. It also provided an empirical relationship relating the heavy element complement of giant planets ($M_Z$) to their host star: $log_{10}(M_Z) = (0.82 \pm 0.08) + (3.40 \pm 0.39)[Fe/H]_{*}$. Applying this relation to HAT-P-13b, which orbits a relatively metal-rich star ([Fe/H]$_* = 0.46\pm 0.07$;~\citet{Torres2012}), we find an estimated heavy element mass of $242_{-160}^{+568}\ M_{\oplus}$, i.e., 84\% of the total mass of HAT-P-13b, a much higher percentage than we determine for the core mass of HAT-P-13b and also a higher percentage than is found for most of the planets considered by~\citet{MillerFortney2011}. This may indicate that the empirical relation cannot be extrapolated to planets around stars with metallicities higher than those of the stars they studied, or that there are additional parameters, such as formation location, that can affect the final core masses for these planets.

\subsection{Future Measurements}

Other systems analogous to the HAT-P-13 system, i.e., systems that allow us to measure the $k_2$ of the inner planet, will be useful for exploring the distribution of core masses over a larger sample of giant planets. In order to exploit the models utilized in this study we require that such a planet (i) be transiting, (ii) have a circularization timescale less than one third of the age of the system, (iii) have an equilibrium eccentricity large enough to be measured with high precision~\citep[Eq. 36 of][]{Mardling2007}, and (iv) have a $\dot{\varpi}_{b_{tid}}$ comparable to or larger than $\dot{\varpi}_{b_{GR}}$~\citep[Eq. 12 of][]{Batygin2011sini}.  Radial velocity observations of the Kepler-424~\citep{Endl2014}, WASP-41~\citep{Neveu-VanMalle2015}, HAT-P-44, HAT-P-45, and HAT-P-46~\citep{Hartman2014} systems indicate that they may have architectures that would make them amenable to this kind of study. We note that many of the hot Jupiters detected by ongoing transit surveys have relatively sparse radial velocity observations, making it difficult to determine whether or not they have a suitable outer companion.~\citet{Knutson2014} find that approximately half of all hot Jupiters have massive long-period companions, suggesting that there is a high probability that future radial velocity campaigns will discover additional systems analogous to HAT-P-13b.

Although the current observations of HAT-P-13 provide an improved estimate of the innermost planet's orbital eccentricity, the uncertainty in this parameter is still the single largest contribution to the uncertainty in the Love number. We therefore conclude that this system could benefit from additional secondary eclipse measurements.

One of our model limitations is the lack of constraint on the metallicity of HAT-P-13b's envelope~\citep[see][]{Kramm2011}. Therefore further atmospheric studies are critical to refine our understanding of HAT-P-13b's structure and composition. Atmospheric circulation models for tidally locked planets suggest that high metallicity atmospheres may have less efficient atmospheric circulation than their lower-metallicity counterparts~\citep{Lewis2010}, which does not appear to be the case for HAT-P-13b based on the atmospheric models we perform. Since HAT-P-13 is currently one of the most metal-rich stars known to host a hot Jupiter, it is intriguing that neither HAT-P-13b's core mass nor its atmosphere suggest significant heavy element enrichment. The HAT-P-13 system will likely provide invaluable leverage when exploring the relationship between host star and planetary metallicity. In addition, full-orbit phase curve observations with Spitzer would also allow us to break degeneracies between the planet's dayside albedo and the efficiency of its atmospheric circulation~\citep[e.g.,][]{Schwartz2015}. The possibility of independently constraining both the core mass and the atmospheric properties of HAT-P-13b makes this planet an ideal target for future observations.

\section{Conclusions}
\label{sec:conclusion}

In this study we present observations of two secondary eclipses of HAT-P-13b centered at 2455326.70818 $\pm$ 0.00406 and 2455355.87672 $\pm$ 0.00226 BJD$_{\rm{UTC}}$.  This corresponds to an error-weighted mean eclipse time that is 17.6 $\pm$ 2.9 minutes minutes earlier (at orbital phase 0.49582 $\pm$ 0.00069) than the predicted time for a circular orbit, indicating that this planet has a non-zero orbital eccentricity. We fit the measured eclipse times simultaneously with the available radial velocity data in order to derive an eccentricity of $e_b = 0.00700 \pm 0.00100$ for this planet, under the assumption that the orbits of HAT-P-13b and HAT-P-13c are coplanar. Using this eccentricity, we calculate a corresponding constraint on the planet's Love number ($k_{2}$). We then use this $k_{2}$ and the measured radius of HAT-P-13b as constraints on interior structure models, which allow us to directly estimate the mass of the planet's core. Moderate mutual inclinations (up to $\sim 10\deg$ between the orbits of HAT-P-13b and HAT-P-13c) do not significantly alter the constraint from $e_b$ on the determination of the core mass. 

We calculate that the core mass of HAT-P-13b is less than 25 $M_{\oplus}$ (9\% of the planet's mass; 68\% confidence interval), with a most likely core mass of 11 $M_{\oplus}$ (4\% of the planet's mass). We also use the secondary eclipse depths to find that the dayside temperature is $1906 \pm 93$ K. Comparing these depths and the dayside temperature to models, we find that it is likely that HAT-P-13b has a strong atmospheric absorber and efficient dayside energy redistribution. 

Obtaining the Love number of HAT-P-13b is crucial to determining its core mass because the presence of a modest core in a Jupiter-mass planet is typically masked by its overlying envelope. The unique opportunity to independently constrain the core mass and atmospheric properties of this hot Jupiter with a modestly sized core makes the HAT-P-13 system an important case study for dynamical constraints on the core masses of gas giant planets.



\acknowledgments

We thank Henry Ngo and Joseph Harrington for insightful discussions. This work is based on observations made with the Spitzer Space Telescope, which is operated by the Jet Propulsion Laboratory, California Institute of Technology, under contract with NASA. This material is based upon work supported by the National Science Foundation Graduate Research Fellowship under grant No. 2014184874. Any opinion, findings, and conclusions or recommendations expressed in this material are those of the authors(s) and do not necessarily reflect the views of the National Science Foundation. This work was based in part on observations obtained at the W. M. Keck Observatory using time granted by the University of Hawaii, the University of California, and the California Institute of Technology.  We thank the observers who contributed to the measurements reported here and acknowledge the efforts of the Keck Observatory staff.  We extend special thanks to those of Hawaiian ancestry on whose sacred mountain of Mauna Kea we are privileged to be guests.


\begin{thebibliography}{}
\expandafter\ifx\csname natexlab\endcsname\relax\def\natexlab#1{#1}\fi

\bibitem[{Agol {et~al.}(2010)Agol, Cowan, Knutson, Deming, Steffen, Henry, \&
  Charbonneau}]{Agol2010}
Agol, E., Cowan, N.~B., Knutson, H.~A., {et~al.} 2010, 18

\bibitem[{Bakos {et~al.}(2009)Bakos, Howard, Noyes, Hartman, Torres, Kovacs,
  Fischer, Latham, Johnson, Marcy, Sasselov, Stefanik, Sipocz, Kovacs,
  Esquerdo, Pal, Lazar, \& Papp}]{Bakos2009}
Bakos, G.~A., Howard, A.~W., Noyes, R.~W., {et~al.} 2009, 11

\bibitem[{Bakos {et~al.}(2015)Bakos, Penev, Bayliss, Hartman, Zhou, Brahm,
  Mancini, de~Val-Borro, Bhatti, Jord\'{a}n, Rabus, Espinoza, Csubry, Howard,
  Fulton, Buchhave, Ciceri, Henning, Schmidt, Isaacson, Noyes, Marcy, Suc,
  Howe, Burrows, L\'{a}z\'{a}r, Papp, \& S\'{a}ri}]{Bakos2015}
Bakos, G.~A., Penev, K., Bayliss, D., {et~al.} 2015, 11

\bibitem[{Ballard {et~al.}(2010)Ballard, Charbonneau, Deming, Knutson,
  Christiansen, Holman, Fabrycky, Seager, \& A'Hearn}]{Ballard2010}
Ballard, S., Charbonneau, D., Deming, D., {et~al.} 2010, Publications of the
  Astronomical Society of the Pacific, 122, 1341

\bibitem[{Baraffe {et~al.}(2008)Baraffe, Chabrier, \& Barman}]{Baraffe2008}
Baraffe, I., Chabrier, G., \& Barman, T. 2008, 1

\bibitem[{Batygin {et~al.}(2009)Batygin, Bodenheimer, \&
  Laughlin}]{Batygin2009}
Batygin, K., Bodenheimer, P., \& Laughlin, G. 2009, 13

\bibitem[{Batygin \& Laughlin(2011)}]{Batygin2011sini}
Batygin, K., \& Laughlin, G. 2011, Astrophysical Journal, 730, 95

\bibitem[{Becker \& Batygin(2013)}]{Becker2013}
Becker, J.~C., \& Batygin, K. 2013, The Astrophysical Journal, 778, 100

\bibitem[{Bodenheimer {et~al.}(2003)Bodenheimer, Laughlin, \&
  Lin}]{Bodenheimer2003}
Bodenheimer, P., Laughlin, G., \& Lin, D. N.~C. 2003, The Astrophysical
  Journal, 592, 555

\bibitem[{Bodenheimer \& Pollack(1986)}]{Bodenheimer1986}
Bodenheimer, P., \& Pollack, J.~B. 1986, 408, 391

\bibitem[{Boss(1997)}]{Boss1997}
Boss, A.~P. 1997, Science, 276, 1836

\bibitem[{Burrows {et~al.}(2008)Burrows, Budaj, \& Hubeny}]{Burrows2008}
Burrows, A., Budaj, J., \& Hubeny, I. 2008, 28

\bibitem[{Burrows {et~al.}(2011)Burrows, Heng, \& Nampaisarn}]{Burrows2011}
Burrows, A., Heng, K., \& Nampaisarn, T. 2011, arXiv:1102.3922

\bibitem[{Burrows {et~al.}(2007)Burrows, Hubeny, Budaj, \&
  Hubbard}]{Burrows2007}
Burrows, A.~S., Hubeny, Budaj, J., \& Hubbard, W.~B. 2007, The Astrophysical
  Journal, 661, 502

\bibitem[{Cowan \& Agol(2011)}]{Cowan2011}
Cowan, N.~B., \& Agol, E. 2011, 12

\bibitem[{Deming {et~al.}(2006)Deming, Harrington, Seager, \&
  Richardson}]{Deming2006}
Deming, D., Harrington, J., Seager, S., \& Richardson, L.~J. 2006, 00, 14

\bibitem[{Deming {et~al.}(2015)Deming, Knutson, Kammer, Burrows, Fortney,
  Carey, Ingalls, \& Fraine}]{Deming2015}
Deming, D., Knutson, H., Kammer, J., {et~al.} 2015

\bibitem[{Endl {et~al.}(2014)Endl, Caldwell, Barclay, Huber, Isaacson,
  Buchhave, Brugamyer, Robertson, Cochran, MacQueen, Havel, Lucas, Howell,
  Fischer, Quintana, \& Ciardi}]{Endl2014}
Endl, M., Caldwell, D.~A., Barclay, T., {et~al.} 2014, The Astrophysical
  Journal, 795, 151

\bibitem[{Fazio {et~al.}(2004)Fazio, Hora, Allen, Ashby, Barmby, Deutsch,
  Huang, Kleiner, Marengo, Megeath, Melnick, Pahre, Patten, Polizotti, Smith,
  Taylor, Wang, Willner, Hoffmann, Pipher, Forrest, Mcmurty, Mccreight,
  Mckelvey, Mcmurray, Koch, Moseley, Arendt, Mentzell, Marx, Losch, Mayman,
  Eichhorn, Krebs, Jhabvala, Gezari, Fixsen, Flores, Shakoorzadeh, Jungo,
  Hakun, Workman, Karpati, Kichak, Whitley, Mann, Tollestrup, Eisenhardt,
  Stern, Gorjian, Bhattacharya, Carey, Nelson, Glaccum, Lacy, Lowrance, Laine,
  Reach, Stauffer, Surace, Wilson, Wright, Hoffman, Domingo, \&
  Cohen}]{Fazio2004}
Fazio, G.~G., Hora, J.~L., Allen, L.~E., {et~al.} 2004, 10

\bibitem[{Foreman-Mackey {et~al.}(2013)Foreman-Mackey, Hogg, Lang, \&
  Goodman}]{ForemanMackey2013}
Foreman-Mackey, D., Hogg, D.~W., Lang, D., \& Goodman, J. 2013, Publications of
  the Astronomical Society of the Pacific, 125, 306

\bibitem[{Fortney {et~al.}(2008)Fortney, Lodders, Marley, \&
  Freedman}]{Fortney2008}
Fortney, J.~J., Lodders, K., Marley, M.~S., \& Freedman, R.~S. 2008, 1419

\bibitem[{Fortney \& Nettelmann(2010)}]{FortneyNettelmann2010}
Fortney, J.~J., \& Nettelmann, N. 2010, Space Science Reviews, 152, 423

\bibitem[{Fortney {et~al.}(2006)Fortney, Saumon, Marley, Lodders, \&
  Freedman}]{Fortney2006}
Fortney, J.~J., Saumon, D., Marley, M.~S., Lodders, K., \& Freedman, R.~S.
  2006, The Astrophysical Journal, 642, 495

\bibitem[{Fulton {et~al.}(2011)Fulton, Shporer, Winn, Holman, P\'{a}l, \&
  Gazak}]{Fulton2011}
Fulton, B.~J., Shporer, A., Winn, J.~N., {et~al.} 2011, 84, arXiv:1105.5599

\bibitem[{Fulton {et~al.}(2013)Fulton, Howard, Winn, Albrecht, Marcy, Crepp,
  Bakos, Johnson, Hartman, Isaacson, Knutson, \& Zhao}]{Fulton2013}
Fulton, B.~J., Howard, A.~W., Winn, J.~N., {et~al.} 2013, The Astrophysical
  Journal, 772, 80

\bibitem[{Guillot {et~al.}(2004)Guillot, Stevenson, Hubbard, \&
  Saumon}]{Guillot2004}
Guillot, T., Stevenson, D.~J., Hubbard, W.~B., \& Saumon, D. 2004, Jupiter: The
  Planet, Satellites and Magnetosphere, 35

\bibitem[{Hartman {et~al.}(2014)Hartman, Bakos, Torres, Kov\'{a}cs, Johnson,
  Howard, Marcy, Latham, Bieryla, Buchhave, Bhatti, B\'{e}ky, Csubry, Penev,
  de~Val-Borro, Noyes, Fischer, Esquerdo, Everett, Szklen\'{a}r, Zhou, Bayliss,
  Shporer, Fulton, Sanchis-Ojeda, Falco, L\'{a}z\'{a}r, Papp, \&
  S\'{a}ri}]{Hartman2014}
Hartman, J.~D., Bakos, G.~A., Torres, G., {et~al.} 2014, The Astronomical
  Journal, 147, 128

\bibitem[{Helled \& Guillot(2013)}]{Helled2013}
Helled, R., \& Guillot, T. 2013, The Astrophysical Journal, 767, 113

\bibitem[{Hubickyj {et~al.}(2005)Hubickyj, Bodenheimer, \&
  Lissauer}]{Hubickyj2005}
Hubickyj, O., Bodenheimer, P., \& Lissauer, J.~J. 2005, Icarus, 179, 415

\bibitem[{Husser {et~al.}(2013)Husser, {Wende-von Berg}, Dreizler, Homeier,
  Reiners, Barman, \& Hauschildt}]{Husser2013}
Husser, T.-O., {Wende-von Berg}, S., Dreizler, S., {et~al.} 2013, Astronomy \&
  Astrophysics, 553, A6

\bibitem[{Ikoma {et~al.}(2006)Ikoma, Guillot, Genda, Takayuki, \&
  Ida}]{Ikoma2006}
Ikoma, M., Guillot, T., Genda, H., Takayuki, T., \& Ida, S. 2006, arXiv:0607212

\bibitem[{Ikoma {et~al.}(2000)Ikoma, Nakazawa, \& Emori}]{Ikoma2000}
Ikoma, M., Nakazawa, K., \& Emori, H. 2000, The Astrophysical Journal, 537,
  1013

\bibitem[{Kammer {et~al.}(2015)Kammer, Knutson, Line, Fortney, Deming, Burrows,
  Cowan, Triaud, Agol, Desert, Fulton, Howard, Laughlin, Lewis, Morley, Moses,
  Showman, \& Todorov}]{Kammer2015}
Kammer, J.~A., Knutson, H.~A., Line, M.~R., {et~al.} 2015, arXiv:1508.00902

\bibitem[{Knutson {et~al.}(2012)Knutson, Lewis, Fortney, Burrows, Showman,
  Cowan, Henry, Agol, Aigrain, Charbonneau, Deming, Langton, \&
  Laughlin}]{Knutson2012}
Knutson, H.~A., Lewis, N., Fortney, J.~J., {et~al.} 2012, 22,
  doi:10.1088/0004-637X/754/1/22

\bibitem[{Knutson {et~al.}(2014)Knutson, Fulton, Montet, Kao, Ngo, Andrew,
  Batygin, Johnson, Crepp, Hinkley, Gaspar, Morton, \& Muirhead}]{Knutson2014}
Knutson, H.~A., Fulton, B.~J., Montet, B.~T., {et~al.} 2014,
  arXiv:arXiv:1312.2954v2

\bibitem[{Kramm {et~al.}(2011)Kramm, Nettelmann, Fortney, Neuh\"{a}user, \&
  Redmer}]{Kramm2011}
Kramm, U., Nettelmann, N., Fortney, J.~J., Neuh\"{a}user, R., \& Redmer, R.
  2011, Astronomy \& Astrophysics, 8

\bibitem[{Leconte \& Chabrier(2012)}]{Leconte2012}
Leconte, J., \& Chabrier, G. 2012, Astronomy \& Astrophysics, 540, 20

\bibitem[{Lewis {et~al.}(2010)Lewis, Showman, Fortney, Marley, Freedman, \&
  Lodders}]{Lewis2010}
Lewis, N.~K., Showman, A.~P., Fortney, J.~J., {et~al.} 2010, 25

\bibitem[{Lewis {et~al.}(2013)Lewis, Knutson, Showman, Cowan, Laughlin,
  Burrows, Deming, Crepp, Mighell, Agol, Bakos, Charbonneau, D\'{e}sert,
  Fischer, Fortney, Hartman, Hinkley, Howard, Johnson, Kao, Langton, \&
  Marcy}]{Lewis2013}
Lewis, N.~K., Knutson, H.~A., Showman, A.~P., {et~al.} 2013, The Astrophysical
  Journal, 766, 95

\bibitem[{Lindegren(2009)}]{Lindegren2009}
Lindegren, L. 2009, Proceedings of the International Astronomical Union, 5, 296

\bibitem[{Loeb(2005)}]{Loeb2005}
Loeb, A. 2005, The Astrophysical Journal, 45

\bibitem[{Love(1911)}]{Love1911}
Love, A. 1911, Cambridge University Press

\bibitem[{Love(1909)}]{Love1909}
Love, A. E.~H. 1909, Proceedings of the Royal Society A: Mathematical, Physical
  and Engineering Sciences, 82, 73

\bibitem[{Mandel \& Agol(2002)}]{Mandel2002}
Mandel, K., \& Agol, E. 2002, The Astrophysical Journal, 580, L171

\bibitem[{Mardling(2007)}]{Mardling2007}
Mardling, R.~A. 2007, Monthly Notices of the Royal Astronomical Society, 382,
  1768

\bibitem[{Mardling(2010)}]{Mardling2010}
---. 2010, Monthly Notices of the Royal Astronomical Society, 407, 1048

\bibitem[{McLaughlin(1924)}]{McLaughlin1924}
McLaughlin, D.~B. 1924, The Astrophysical Journal, 60, 22

\bibitem[{Miller \& Fortney(2011)}]{MillerFortney2011}
Miller, N., \& Fortney, J.~J. 2011, 13

\bibitem[{Mizuno(1980)}]{Mizuno1980}
Mizuno, H. 1980, Progress of Theoretical Physics, 64, 544

\bibitem[{Nesvorn\'{y}(2009)}]{Nesvorny2009}
Nesvorn\'{y}, D. 2009, The Astrophysical Journal, 701, 1116

\bibitem[{Nettelmann {et~al.}(2012)Nettelmann, Becker, Holst, \&
  Redmer}]{Nettelmann2012}
Nettelmann, N., Becker, A., Holst, B., \& Redmer, R. 2012, The Astrophysical
  Journal, 750, 52

\bibitem[{Neveu-VanMalle {et~al.}(2015)Neveu-VanMalle, Queloz, Anderson, Brown,
  {Collier Cameron}, Delrez, Diaz, Gillon, Hellier, Jehin, Lister, Pepe, Rojo,
  S\'{e}gransan, Triaud, \& Turner}]{Neveu-VanMalle2015}
Neveu-VanMalle, M., Queloz, D., Anderson, D., {et~al.} 2015, 41, 1

\bibitem[{Paxton {et~al.}(2011)Paxton, Bildsten, Dotter, Herwig, Lesaffre, \&
  Timmes}]{Paxton2011}
Paxton, B., Bildsten, L., Dotter, A., {et~al.} 2011, 3, 110

\bibitem[{Payne \& Ford(2011)}]{PayneFord2011}
Payne, M.~J., \& Ford, E.~B. 2011, 98, arXiv:1103.0199

\bibitem[{Perez-Becker \& Showman(2013)}]{Perez-Becker2013}
Perez-Becker, D., \& Showman, A. 2013, arXiv preprint arXiv:1306.4673, 1

\bibitem[{Pollack {et~al.}(1995)Pollack, Hubickyj, Bodenheimer, Lissauer,
  Podolak, \& Greenzweig}]{Pollack1996}
Pollack, J.~B., Hubickyj, O., Bodenheimer, P., {et~al.} 1995, 85, 62

\bibitem[{Pont {et~al.}(2006)Pont, Zucker, \& Queloz}]{Pont2006}
Pont, F., Zucker, S., \& Queloz, D. 2006, Monthly Notices of the Royal
  Astronomical Society, 373, 231

\bibitem[{Rafikov(2006)}]{Rafikov2006}
Rafikov, R. 2006, 19

\bibitem[{Ragozzine \& Wolf(2008)}]{Ragozzine2008}
Ragozzine, D., \& Wolf, A.~S. 2008, 19

\bibitem[{Rossiter(1924)}]{Rossiter1924}
Rossiter, R.~A. 1924, The Astrophysical Journal, 60, 15

\bibitem[{Safronov(1969)}]{Safronov1969}
Safronov, V. 1969, Nauka, Moscow. English translation: NASA TTF-677. 1972.

\bibitem[{Sato {et~al.}(2005)Sato, Fischer, Henry, Laughlin, Butler, Marcy,
  Vogt, Bodenheimer, Ida, Toyota, Wolf, Valenti, Boyd, Johnson, Wright, Ammons,
  Robinson, Strader, McCarthy, Tah, \& Minniti}]{Sato2005}
Sato, B., Fischer, D.~A., Henry, G.~W., {et~al.} 2005, 1

\bibitem[{Saumon {et~al.}(1995)Saumon, Chabrier, \& van Horn}]{Saumon1995}
Saumon, D., Chabrier, G., \& van Horn, H.~M. 1995, The Astrophysical Journal
  Supplement Series, 99, 713

\bibitem[{Schwartz \& Cowan(2015)}]{Schwartz2015}
Schwartz, J.~C., \& Cowan, N.~B. 2015, Monthly Notices of the Royal
  Astronomical Society, 449, 4192

\bibitem[{Southworth(2010)}]{Southworth2010}
Southworth, J. 2010, Monthly Notices of the Royal Astronomical Society, 408,
  1689

\bibitem[{Southworth {et~al.}(2012)Southworth, Bruni, Mancini, \&
  Gregorio}]{Southworth2012}
Southworth, J., Bruni, I., Mancini, L., \& Gregorio, J. 2012, Monthly Notices
  of the Royal Astronomical Society, 420, 2580

\bibitem[{Sterne(1939)}]{Sterne1939}
Sterne, T.~E. 1939, Monthly Notices of the Royal Astronomical Society,
  99, 451

\bibitem[{Stevenson(1982)}]{Stevenson1982}
Stevenson, D.~J. 1982, Planetary and Space Science, 30, 755

\bibitem[{Torres {et~al.}(2012)Torres, Fischer, Sozzetti, Buchhave, Winn,
  Holman, \& Carter}]{Torres2012}
Torres, G., Fischer, D.~A., Sozzetti, A., {et~al.} 2012, arXiv:1208.1268

\bibitem[{Torres {et~al.}(2007)Torres, Bakos, Kovacs, Latham, Fernandez, Noyes,
  Esquerdo, Sozzetti, Fischer, Butler, Marcy, Stefanik, Sasselov, Lazar, Papp,
  \& Sari}]{Torres2007}
Torres, G., Bakos, G.~A., Kovacs, G., {et~al.} 2007, 1

\bibitem[{Winn {et~al.}(2007)Winn, Holman, Bakos, P\'{a}l, Johnson, Williams,
  Shporer, Mazeh, Fernandez, Latham, \& Gillon}]{Winn2007}
Winn, J.~N., Holman, M.~J., Bakos, G.~A., {et~al.} 2007, The Astronomical
  Journal, 134, 1707

\bibitem[{Winn {et~al.}(2008)Winn, Holman, Torres, McCullough, Johns‐Krull,
  Latham, Shporer, Mazeh, Garcia‐Melendo, Foote, Esquerdo, \&
  Everett}]{Winn2008}
Winn, J.~N., Holman, M.~J., Torres, G., {et~al.} 2008, The Astrophysical
  Journal, 683, 1076

\bibitem[{Winn {et~al.}(2010)Winn, Johnson, Howard, Marcy, Bakos, Hartman,
  Torres, Albrecht, \& Narita}]{Winn2010}
Winn, J.~N., Johnson, J.~A., Howard, A.~W., {et~al.} 2010, 20

\bibitem[{Wong {et~al.}(2014)Wong, Knutson, Cowan, Lewis, Agol, Burrows,
  Deming, Fortney, Fulton, Langton, Laughlin, \& Showman}]{Wong2014}
Wong, I., Knutson, H.~A., Cowan, N.~B., {et~al.} 2014, 13

\end{thebibliography}
\end{document}